\documentclass[12pt]{iopart}
\usepackage{iopams}

 \newcommand{\sech}{\rm \ sech \,}
 \newcommand{\cosech}{\rm \ cosech \,}
\newcommand{\cosec}{\rm \ cosec \,}
 \newcommand{\cotan}{\rm \ cotan \,}
\newcommand{\cotanh}{\rm \ cotanh \,}

\begin{document}
\title[Orthogonal polynomials defined by hypergeometric type 
equations]{Systems of orthogonal polynomials defined by 
hypergeometric type equations with application to quantum mechanics}
\author{Nicolae Cotfas}
\address{Faculty of Physics, University of Bucharest,
PO Box 76-54, Postal Office 76, Bucharest, Romania,\\ 
E-mail address: ncotfas@yahoo.com\ \ \ 
http://fpcm5.fizica.unibuc.ro/\~{ }ncotfas}
\jl{1}
\begin{abstract}
A hypergeometric type equation satisfying certain conditions
defines either a finite or an 
infinite system of orthogonal polynomials. We present in a unified and
explicit way all these systems of orthogonal polynomials, the associated
special functions and the corresponding raising/lowering operators. The
considered equations are directly related to some Schr\" odinger type
equations (P\" oschl-Teller, Scarf, Morse, etc), and the defined special 
functions are related to the corresponding bound-state eigenfunctions. 
\end{abstract}
\maketitle
\newpage

\section{Introduction}

Many problems in quantum mechanics and
mathematical physics lead to equations of the type
\begin{equation}\label{hypeq}
\sigma (s)y''(s)+\tau (s)y'(s)+\lambda y(s)=0 \label{eq}
\end{equation}
where $\sigma (s)$ and $\tau (s)$ are polynomials of at most second
and first degree, respectively, and $\lambda $ is a constant. 
These equations are usually called {\em equations of hypergeometric
type} \cite{NUS}, and each of them can be reduced to the self-adjoint form 
\begin{equation}
[\sigma (s)\varrho (s)y'(s)]'+\lambda \varrho (s)y(s)=0 
\end{equation}
by choosing a function $\varrho $ such that 
$[\sigma (s)\varrho (s)]'=\tau (s)\varrho (s)$.

The equation (\ref{hypeq}) is usually considered on an interval $(a,b)$,
chosen such that 
\begin{equation}\begin{array}{r}
\sigma (s)>0\qquad {\rm for\ all}\quad s\in (a,b)\\
\varrho (s)>0\qquad {\rm for\ all}\quad s\in (a,b)\\
\lim_{s\rightarrow a}\sigma (s)\varrho (s)
=\lim_{s\rightarrow b}\sigma (s)\varrho (s)=0.
\end{array}
\end{equation}
Since the form of the equation (\ref{eq}) is invariant under an affine 
change of variable, it is sufficient to analyse the cases
presented in table 1.
Some restrictions are to be imposed to $\alpha $, $\beta $ in
order the interval $(a,b)$ to exist.\\

{\bf Table 1.} The main particular cases.\\[2mm]

\begin{tabular}{cclll}
\hline
$\sigma (s)$ & $\tau (s)$  & $\varrho (s)$ & $\alpha ,\beta $ &  
$(a,b) $\\
\hline
$1$ & $\alpha s+\beta $ & $\e^{\alpha s^2/2+\beta s}$ & $\alpha <0$
& $(-\infty , \infty )$\\
$s$ & $\alpha s+\beta $ & $s^{\beta -1} \e^{\alpha s}$ & 
$\alpha <0$, $\beta >0$& $(0,\infty )$\\ 
$1-s^2$  & $\alpha s+\beta $ & $(1+s)^{-(\alpha -\beta )/2-1}
(1-s)^{-(\alpha +\beta )/2-1}$ & 
$\alpha <\beta <-\alpha $ & $(-1,1)$\\
$s^2-1$  & $\alpha s+\beta $ & $(s+1)^{(\alpha -\beta )/2-1}
(s-1)^{(\alpha +\beta )/2-1}$ &
$-\beta <\alpha <0$ & $(1,\infty )$\\
$s^2$  & $\alpha s+\beta $ & $s^{\alpha -2}\e^{-\beta /s}$ & $\alpha <0$, $\beta >0$ &
$(0,\infty )$\\
$s^2+1$  & $\alpha s+\beta $ & $(1+s^2)^{\alpha /2-1}\e^{\beta \arctan s}$ & 
$\alpha <0$ & $(-\infty ,\infty )$\\
\hline
\end{tabular}

\vspace{5mm}

\indent 
Our main purpose is to present a unified view on the systems of orthogonal 
polynomials defined by equation (\ref{hypeq}) in the  cases
presented in table 1. We analyse the associated special functions, the
corresponding raising/lowering operators, and present some applications to
quantum mechanics. Some of our results and proofs are extended versions of some 
results and proofs presented in \cite{NUS}. The general form of the 
raising/lowering operators has been previously obtained by Jafarizadeh and Fakhri 
by using a different method \cite{JF}. Our results concerning the relation with 
classical polynomials obtained in a direct way are in agreement with those based on 
recursion relations presented in \cite{CKS,D}.

\section{Polynomials of hypergeometric type}

It is well-known \cite{NUS} that for $\lambda =\lambda _l$, where
\begin{equation}
\lambda _l=-\frac{\sigma ''(s)}{2}l(l-1)-\tau '(s)l\qquad  l\in \mathbb{N}
\end{equation}
the equation (\ref{hypeq}) admits a polynomial solution 
$\Phi _l=\Phi _l^{(\alpha ,\beta )}$ of at most $l$ degree
\begin{equation} \label{eq3}
\sigma (s) \Phi _l ''+\tau (s) \Phi _l '+\lambda _l\Phi _l=0.
\end{equation}
If the degree of the polynomial $\Phi _l$ is $l$ then it satisfies the
Rodrigues formula
\begin{equation}
\Phi _l(s)=\frac{B_l}{\varrho (s)}[\sigma ^l(s)\varrho (s)]^{(l)}
\end{equation}
where $B_l$ is a constant. We do not impose any normalizing condition.
Our polynomials $\Phi _l$ are defined only up to a multiplicative constant. 
One can remark that
\begin{equation}\label{gamma1}
\lim_{s\rightarrow a}\sigma (s)\varrho (s)s^\gamma 
=\lim_{s\rightarrow b}\sigma (s)\varrho (s)s^\gamma =0\qquad 
{\rm for }\quad \gamma \in [0,\infty ) 
\end{equation}
in the case $\sigma (s)\in \{ 1,\ s,\ 1-s^2\}$, and 
\begin{equation}\label{gamma2}
\lim_{s\rightarrow a}\sigma (s)\varrho (s)s^\gamma 
=\lim_{s\rightarrow b}\sigma (s)\varrho (s)s^\gamma =0\qquad 
{\rm for }\quad \gamma \in [0,-\alpha )
\end{equation}
in the case $\sigma (s)\in \{ s^2-1,\ s^2,\ s^2+1\}$.
Let 
\begin{equation}
\nu =\left\{ \begin{array}{lcl}
\infty & {\rm for} & \sigma (s)\in \{ 1,\ s,\ 1-s^2\}\\[2mm]
\frac{1-\alpha }{2} & {\mbox{}\quad \rm for \quad \mbox{}} & 
\sigma (s)\in \{ s^2-1,\ s^2,\ s^2+1\} .\\[3mm]
\end{array}\right.
\end{equation}
{\bf Theorem 1.} {\it \\
a) $\{\Phi _l\ |\ l<\nu \}$ is a system of polynomials orthogonal 
with weight function $\varrho (s)$ in $(a,b)$. \\[2mm]
b) \ $\Phi _l$ is a
polynomial of degree $l$ for any $l<\nu $.\\[2mm]
c) The function $\Phi _l(s)\sqrt{\varrho (s)}$
is square integrable on $(a,b)$ for any $l<\nu $.\\[2mm]
d) A three term recurrence relation
\[ s\Phi _l(s)=\alpha _l \Phi _{l+1}(s)+\beta _l\Phi _l(s)
+\gamma _l\Phi _{l-1}(s) \]
is satisfied for $1<l+1<\nu $.\\[2mm]
e) The zeros of $\Phi _l$ are simple and lie in the interval $(a,b)$,
for any $l<\nu $. }\\[3mm]
{\bf Proof.} 
a) Let $l,\ k\in \mathbb{N}$ with $0\leq l<k<\nu $. From the relations 
\[ [\sigma (s) \varrho (s)\Phi '_l]'+\lambda _l\varrho (s)\Phi _l =0\qquad 
[\sigma (s) \varrho (s)\Phi '_k]'+\lambda _k\varrho (s)\Phi _k =0 \]
we get
\[ (\lambda _l-\lambda _k)\Phi _l(s)\Phi _k(s)\varrho (s)
=\frac{d}{ds}\{ \sigma (s)\varrho (s) 
[\Phi _l(s)\Phi '_k(s)-\Phi _k(s)\Phi '_l(s)]\} .\]
Since the Wronskian $W[\Phi _l(s),\Phi _k(s)]
=\Phi _l(s)\Phi '_k(s)-\Phi _k(s)\Phi '_l(s)$ is a polynomial of 
at most $l+k-1$ degree, from (\ref{gamma1}) and (\ref{gamma2}) it follows that
\[ (\lambda _l-\lambda _k)\int_a^b\Phi _l(s)\Phi _k(s)\varrho (s)ds\]
\[ =\lim_{s\rightarrow b}\sigma (s)\varrho (s)W[\Phi _l(s),\Phi _k(s)]
-\lim_{s\rightarrow a}\sigma (s)\varrho (s)W[\Phi _l(s),\Phi _k(s)]=0.\]
We have $\lambda _l-\lambda _k\not=0$ since the function 
$l\mapsto \lambda _l$ is strictly
increasing on  $\{ l\in \mathbb{N}\ |\ l<\nu \}$.\\
b)
Each $\Phi _l$ is a polynomial of at most $l$ degree and
the polynomials $\{ \Phi _l\ |\ 0\leq l<\nu \}$ are 
linearly independent. This is possible only if $\Phi _l$ is a
polynomial of degree $l$ for any $l<\nu $.\\
c) In the case $\sigma (s)\in \{ s^2-1,\ s^2,\ s^2+1\}$ we have
$l<(1-\alpha )/2$, that is, $2l-1<-\alpha $. Therefore there exists 
$\varepsilon >0$ such that
$1+\varepsilon +2l-2=2l-1+\varepsilon <-\alpha $, and hence
\[  \lim_{s\rightarrow \infty } s^{1+\varepsilon }|\Phi _l(s)|^2\varrho (s)
=\lim_{s\rightarrow \infty } \frac{s^{1+\varepsilon }(\Phi _l(s))^2}
{\sigma (s)}\sigma (s)\varrho (s)=0.\]
The convergence of the integral 
\[ \int_a^b|\Phi _l(s)|^2\varrho (s)ds \]
follows from the convergence 
of the integral $\int_1^\infty s^{-1-\varepsilon }ds$. 
In the case $\sigma (s)=1-s^2$, from $\alpha <\beta <-\alpha $ we get
$-(\alpha -\beta )/2-1>-1$ and $-(\alpha +\beta /2-1>-1$. \\
d,e) See \cite{NUS}.\\[3mm]
{\bf Theorem 2.} {\it Up to a multiplicative constant
\begin{equation}\label{classical}
\fl \Phi _l^{(\alpha ,\beta )}(s)=\left\{ \begin{array}{lcl}
H_l\left(\sqrt{\frac{-\alpha }{2}}\, s-\frac{\beta }{\sqrt{-2\alpha }}\right)  
& {\mbox{}\quad {\rm in\ the\ case}\quad \mbox{}} & \sigma (s)=1\\[2mm]
L_l^{\beta -1}(-\alpha s)  & {\rm in\ the\ case} & \sigma (s)=s\\[2mm]
P_l^{(-(\alpha +\beta )/2-1,\ (-\alpha +\beta )/2-1)}(s)  & {\rm in\ the\ case} & \sigma (s)=1-s^2\\[2mm]
P_l^{((\alpha -\beta )/2-1,\ (\alpha +\beta )/2-1)}(-s)  & {\rm in\ the\ case } & \sigma (s)=s^2-1\\[2mm]
\left(\frac{s}{\beta }\right)^lL_l^{1-\alpha -2l}\left(\frac{\beta }{s}\right) 
& {\rm in\ the\ case} & \sigma (s)=s^2\\[2mm]
{\rm i}^lP_l^{((\alpha +{\rm i}\beta )/2-1,\ (\alpha -{\rm i}\beta )/2-1)}({\rm i}s) 
& {\rm in\ the\ case} & \sigma (s)=s^2+1
\end{array} \right.
\end{equation}
where $H_n$, $L_n^p $ and $P_n^{(p,q)}$ are the Hermite,
Laguerre and Jacobi polynomials, respectively}.\\[3mm]
{\bf Proof.} In the case $\sigma (s)=s^2$ the function $\Phi _l^{(\alpha ,\beta )}(s)$
satisfies the equation
\[ s^2y''+(\alpha s+\beta )y'+[-l(l-1)-\alpha l]y=0. \]
If we denote $t=\beta /s$ then the polynomial $u(t)=t^ly(\beta /t)$ satisfies the equation
\[ tu''+(-\alpha +2-2l-t)u'+lu=0 \]
that is, the equation whose polynomial solution is $L_l^{1-\alpha -2l}(s)$ (up to
a multiplicative constant).
In a similar way one can analyse the other cases.

\section{Special functions of hypergeometric type}

Let $l\in \mathbb{N}$, $l<\nu $, and let $m\in \{ 0,1,...,l\}$.
By differentiating the equation (\ref{eq3}) $m$ times we obtain 
the equation satisfied by the polynomials 
$\varphi _{l,m}=\Phi _l^{(m)}$, namely
\begin{equation}\label{varphi}
 \sigma (s)\varphi ''_{l,m}
+[\tau (s)+m\sigma '(s)]\varphi '_{l,m}
+(\lambda _l-\lambda _m) \varphi _{l,m}=0.
\end{equation}
This is an equation of hypergeometric type, and we can write 
it in the self-adjoint form 
\begin{equation}
[\sigma (s) \varrho _m(s)\varphi '_{l,m}]'
+(\lambda _l-\lambda _m)\varrho _m(s)\varphi _{l,m} =0
\end{equation}
by using the function  $\varrho _m(s)=\sigma ^m(s)\varrho (s)$.

The functions 
\begin{equation}\label{def}
\Phi _{l,m}(s)=\kappa ^m(s)\Phi _l^{(m)}(s) 
\end{equation}
where $l\in \mathbb{N}$, $l<\nu $, $m\in \{ 0,1,...,l\}$
and $\kappa (s)=\sqrt{\sigma (s)}$,  
are called the {\em associated special functions}.
The equation (\ref{varphi}) multiplied by $\kappa ^m(s)$ 
can be written as
\begin{equation}\label{Hm}
H_m \Phi _{l,m}=\lambda _l\Phi _{l,m}
\end{equation}
where $H_m$ is the differential operator
\begin{eqnarray}\label{defHm}
H_m= & - & \sigma (s) \frac{d^2}{ds^2}-\tau (s) \frac{d}{ds}
+\frac{m(m-2)}{4}\frac{{\sigma '}^2(s)}{\sigma (s)} 
\nonumber \\ & + & 
\frac{m\tau (s)}{2}\frac{\sigma '(s)}{\sigma (s)}
-\frac{1}{2}m(m-2)\sigma ''(s)-m\tau '(s) .
\end{eqnarray}
{\bf Theorem 3.} {\it 
a) For each $m<\nu $, the functions
$\Phi _{l,m}$ with $m\leq l<\nu $ 
are orthogonal with weight function $\varrho (s)$ in $(a,b)$.\\[2mm] 
b) $\Phi _{l,m}(s)\sqrt{\varrho (s)}$
is square integrable on $(a,b)$ for $0\leq m\leq l<\nu $.\\[2mm]
c) The three term recurrence relation
\begin{eqnarray}
 \fl \Phi _{l,m+1}(s) + \left( \frac{\tau (s)}{\kappa (s)}
+2(m-1)\kappa '(s)\right)\Phi _{l,m}(s) 
+(\lambda _l-\lambda _{m-1}) \Phi _{l,m-1}(s)=0 \label{rec}
\end{eqnarray}
is satisfied for any $l<\nu $ and any $m\in \{ 1,2,...,l-1\}$.
In addition, we have}
\begin{equation}\label{rec1}
\left( \frac{\tau (s)}{\kappa (s)}+
2(l-1)\kappa '(s)\right) \Phi _{l,l}(s)
+(\lambda _l-\lambda _{l-1})\Phi _{l,l-1}(s)=0.
\end{equation}
{\bf Proof.} 
a) In the case $\sigma (s)\in \{ s^2-1,\ s^2,\ s^2+1\}$,
for each $k<-\alpha -2m$ we have
\begin{equation}
\lim_{s\rightarrow a}\sigma (s)\varrho _m(s)s^k
=\lim_{s\rightarrow b}\sigma (s)\varrho _m(s)s^k=0.
\end{equation}
Since the equation (\ref{varphi}) is of hypergeometric type, 
$\varphi _{l,m}$ is a polynomial of degree $l-m$, 
and $\varphi _{k,m}$ is a polynomial of degree $k-m$, 
from theorem 1 applied to (\ref{varphi}) it follows that
\[ \int_a^b\varphi _{l,m}(s)\varphi _{k,m}(s)\varrho _m(s)ds=0 \]
for $l\not=k$ with $(l-m)+(k-m)<-\alpha -2m$, that is, $l+k<-\alpha $.
But $\varrho _m(s)=\sigma ^m(s)\varrho (s)$, and  hence
\[ \int_a^b\varphi _{l,m}(s)\varphi _{k,m}(s)\varrho _m(s)ds
=\int_a^b\Phi _{l,m}(s)\Phi _{k,m}(s)\varrho (s)ds.\]
b) In the case $\sigma (s)\in \{ s^2-1,\ s^2,\ s^2+1\}$,
for $\varepsilon >0$ chosen such that
$2\nu +\varepsilon <1-\alpha $ we have 
$1+\varepsilon +2(m-1)+2(l-m)\leq 1+\varepsilon +2\nu -2<-\alpha $, whence
\[ \lim_{s\rightarrow \infty }s^{1+\varepsilon }(\Phi _{l,m}(s))^2\varrho (s)
=\lim_{s\rightarrow \infty }s^{1+\varepsilon }
\sigma ^{m-1}(s)(\Phi _l^{(m)}(s))^2\sigma (s)\varrho (s)=0.\]
The convergence of the integral 
\[ \int_a^b|\Phi _{l,m}(s)|^2\varrho (s)ds \]
follows from the 
convergence of the integral $\int_1^\infty s^{-1-\varepsilon }ds$ .\\
c)
By differentiating (\ref{eq3}) $m-1$ times we obtain
\[   \sigma (s)\Phi _l^{(m+1)}(s)+(m-1)\sigma '(s)\Phi _l^{(m)}(s)+
\frac{(m-1)(m-2)}{2}\sigma ''(s)\Phi _l^{(m-1)}(s)\]
\[ +\tau (s)\Phi _l^{(m)}+(m-1)\tau '(s)\Phi _l^{(m-1)}(s)
+\lambda _l\Phi _l^{(m-1)}(s)=0.\]
If we multiply this relation by $\kappa ^{m-1}(s)$ then we get 
(\ref{rec}) for $m\in \{ 1,2,...,l-1\}$, and (\ref{rec1}) for $m=l$.
$\qquad   $\\[2mm]

\section{Raising and lowering operators}

For any $l\in \mathbb{N}$, $l<\nu $ and any $m\in \{ 0,1,...,l-1\}$, 
by differentiating (\ref{def}), we obtain
\[ \Phi '_{l,m}(s)=m\kappa ^{m-1}(s)\kappa '(s)\Phi _l^{(m)}
+\kappa ^m(s)\Phi _l^{(m+1)}(s) \]
that is, the relation 
\[ \Phi '_{l,m}(s)=m\frac{\kappa '(s)}{\kappa (s)}\Phi _{l,m}(s)+
\frac{1}{\kappa (s)}\Phi_{l,m+1}(s) \]
which can be written as
\begin{equation}
\left(\kappa (s)\frac{d}{ds}-
m\kappa '(s)\right) \Phi _{l,m}(s)=\Phi _{l,m+1}(s).\label{raising}
\end{equation}

If $m\in \{ 1,2,...,l-1\}$ then by substituting (\ref{raising}) into
(\ref{rec}) we get 
\[   \left( \kappa (s)\frac{d}{ds}+\frac{\tau (s)}{\kappa (s)}
+(m-2)\kappa '(s)\right)\Phi _{l,m}(s) 
+(\lambda _l-\lambda _{m-1}) \Phi _{l,m-1}(s)=0 \]
that is,
\begin{equation}
  \left(-\kappa (s)\frac{d}{ds}-
   \frac{\tau (s)}{\kappa (s)}-(m-1)\kappa '(s)\right)
\Phi _{l,m+1}(s)=(\lambda _l-\lambda _m)\Phi _{l,m}(s).\label{lowering}
\end{equation}
for all $m\in \{ 0,1,...,l-2\}$. From (\ref{rec1}) it follows that this 
relation is also satisfied for $m=l-1$.

The relations (\ref{raising})  and (\ref{lowering}) suggest us to consider 
the first order differential operators
\begin{equation}
  A_m=\kappa (s)\frac{d}{ds}-m\kappa '(s)\qquad
A_m^+=-\kappa (s)\frac{d}{ds}-\frac{\tau (s)}{\kappa (s)}-(m-1)\kappa '(s)
\end{equation}
for $m+1<\nu $.\\[3mm]

\noindent{\bf Theorem 4.} {\it We have }
 \begin{equation}\label{am+}
\fl  a)\quad A_m\Phi _{l,m}=\Phi _{l,m+1}\qquad 
A_m^+\Phi _{l,m+1}=(\lambda _l-\lambda _m)\Phi _{l,m}\qquad for \quad 0\leq m<l< \nu .
\end{equation}
\begin{equation}\label{philm}
\fl b)\quad \Phi _{l,m}=
\frac{A_m^+ }{\lambda _l-\lambda _m}
\frac{A_{m+1}^+ }{\lambda _l-\lambda _{m+1}}...
\frac{A_{l-2}^+ }{\lambda _l-\lambda _{l-2}}
\frac{A_{l-1}^+ }{\lambda _l-\lambda _{l-1}}\Phi _{l,l}\qquad for \quad  0\leq m<l< \nu .
\end{equation}
\begin{equation} 
\fl c)\quad ||\Phi _{l,m+1}||
=\sqrt{\lambda _l-\lambda _m}\, ||\Phi _{l,m}||\qquad  
for\quad  0\leq m<l< \nu .
\end{equation}
\begin{equation}\label{fact}
\fl d)\quad H_m-\lambda _m=A_m^+A_m\qquad H_{m+1}-\lambda _m=A_mA_m^+ \qquad for\quad m+1<\nu 
\end{equation} 
\begin{equation}\label{interw}
\fl e)\quad H_mA_m^+=A_m^+H_{m+1}\qquad A_mH_m=H_{m+1}A_m\qquad for \quad m+1<\nu .
\end{equation}
{\bf Proof.} 
a)
These relations coincide to (\ref{raising})  and (\ref{lowering}),
respectively.\\
b)
Since the function $k\mapsto \lambda _k$ is strictly increasing on
$\{ k\in \mathbb{N}\ |\ k< \nu \}$ we have $\lambda _l-\lambda _k\not=0$ for $k\not= l $.
The formula (\ref{philm}) follows from (\ref{am+}).\\
c)
Since $\sigma ^m(s)\Phi _l^{(m)}(s)\Phi _k^{(m+1)}(s)$ is a polynomial
of degree $l+k-1$, from (\ref{gamma1}) and (\ref{gamma2}) we get 
\[   \langle A_m\Phi _{l,m},\Phi _{k,m+1} \rangle =
\int_a^b[\kappa (s)\Phi '_{l,m}(s) -
m\kappa '(s)\Phi _{l,m}(s)]\Phi _{k,m+1}(s)\varrho (s)ds\]
\[   =\kappa (s)\Phi _{l,m}(s)\Phi _{k,m+1}(s)\varrho (s)|_a^b
-\int_a^b\Phi _{l,m}(s)[\kappa (s)\Phi '_{k,m+1}(s)\varrho (s)\]
\[   +\kappa (s)\Phi _{k,m+1}(s)\varrho '(s)
+(m+1)\kappa '(s)\Phi _{k,m+1}(s)\varrho (s)]ds\]
\[   =\sigma (s)\varrho (s)\sigma ^m(s)
\Phi _l^{(m)}(s)\Phi _k^{(m+1)}(s)|_a^b
+\int_a^b \Phi _{l,m} (s)(A_m^+ \Phi _{k,m+1})(s)\varrho (s)ds\]
\[ =\langle \Phi _{l,m} ,A_m^+\Phi _{k,m+1} \rangle \]
whence
\[ \fl ||\Phi _{l,m+1}||^2
=\langle \Phi _{l,m+1},\Phi _{l,m+1}\rangle
=\langle A_m\Phi _{l,m},\Phi _{l,m+1}\rangle 
 =  \langle \Phi _{l,m},A_m^+\Phi _{l,m+1}\rangle
=(\lambda _l-\lambda _m)||\Phi _{l,m}||^2 .\]
d,e) These relations can be proved by direct computation.

\section{Application to Schr\" odinger type operators}

If we use in equation (\ref{Hm}) a change of variable 
$(a',b')\longrightarrow (a,b):x\mapsto s(x)$ 
such that $ds/dx=\kappa (s(x))$ or $ds/dx=-\kappa (s(x))$ and
define the new functions 
\begin{equation}
\Psi _{l,m}(x)=\sqrt{\kappa (s(x))\, \varrho (s(x))}\, \Phi _{l,m}(s(x))
\end{equation}
then we get an equation of Schr\" odinger type \cite{IH}
\begin{equation}
-\frac{d^2}{dx^2}\Psi _{l,m}(x)+V_m(x)\Psi _{l,m}(x)
=\lambda _l\Psi _{l,m}(x) .\label{Schrod}
\end{equation}
Since
\[  \fl \int_{a'}^{b'}|\Psi _{l,m}(x)|^2dx=
\int_{a'}^{b'}|\Phi _{l,m}(s(x))|^2\varrho (s(x))\frac{d}{dx}s(x)dx
=\int_a^b|\Phi _{l,m}(s)|^2\varrho (s)ds \]
\[   \int_{a'}^{b'}\Psi _{l,m}(x)\Psi _{k,m}(x)dx
=\int_a^b\Phi _{l,m}(s)\Phi _{k,m}(s)\varrho (s)ds \]
the functions $\Psi _{l,m}(x)$ with $0\leq m\leq l<\nu $ 
are square integrable on $(a',b')$ and orthogonal.

If $ds/dx=\kappa (s(x))$ then the operators corresponding to $A_m$ and $A_m^+ $ are
\begin{equation}\label{tildeA+}
 \begin{array}{l}
{\mathcal A}_m=[\kappa (s)\varrho (s)]^{1/2}A_m
[\kappa (s)\varrho (s)]^{-1/2}|_{s=s(x)}
=\frac{d}{dx}+W_m(x)\\[2mm]
{\mathcal A}_m^+ =[\kappa (s)\varrho (s)]^{1/2}A_m^+ 
[\kappa (s)\varrho (s)]^{-1/2}|_{s=s(x)}
=-\frac{d}{dx}+W_m(x)
\end{array}
\end{equation}
where the {\em superpotential} $W_m(x)$ is given by the formula \cite{JF}
\begin{equation}\label{Wm}
W_m(x)=-\frac{\tau (s(x))}{2\kappa (s(x))}
-\frac{2m-1}{2\kappa (s(x))}\frac{d}{dx}\kappa (s(x))\, .
\end{equation}

From (\ref{am+}) and (\ref{fact}) we get
\begin{equation}
  {\mathcal A}_m\Psi _{l,m}(x)=\Psi _{l,m+1}(x) \qquad 
{\mathcal A}_m^+ \Psi _{l,m+1}(x)=(\lambda _l-\lambda _m)\Psi _{l,m}(x)
\end{equation}
and
\begin{equation}
  ({\mathcal A}_m^+ {\mathcal A}_m+\lambda _m)\Psi _{l,m}=
\lambda _l\Psi _{l,m}\qquad 
({\mathcal A}_m{\mathcal A}_m^+ +\lambda _m)\Psi _{l,m+1}=
\lambda _l\Psi _{l,m+1}
\end{equation}
whence
\begin{equation}\label{factor}
 \fl -\frac{d^2}{dx^2}+V_m(x)-\lambda _m={\mathcal A}_m^+ {\mathcal A}_m\qquad
-\frac{d^2}{dx^2}+V_{m+1}(x)-\lambda _m={\mathcal A}_m{\mathcal A}_m^+ 
\end{equation}
and
\begin{equation}\label{Vm}
 \fl V_m(x)-\lambda _m=W_m^2(x)-\dot W_m(x)\qquad
V_{m+1}(x)-\lambda _m=W_m^2(x)+\dot W_m(x)
\end{equation}
where the dot sign means derivative with respect to $x$.

Since ${\mathcal A}_m\Psi _{m,m}=0$, from  (\ref{tildeA+}) and 
(\ref{factor}) we get
\begin{equation}
  \dot \Psi _{m,m}+W_m(x)\Psi _{m,m}=0\qquad 
-\ddot \Psi _{m,m}+(V_m(x)-\lambda _m)\Psi _{m,m}=0
\end{equation}
whence
\begin{equation}\label{WV}
W_m(x)=-\frac{\dot \Psi _{m,m}(x)}{\Psi _{m,m}(x)}\qquad \qquad
V_m(x)=\frac{\ddot \Psi _{m,m}(x)}{\Psi _{m,m}(x)}+\lambda _m .
\end{equation}
For each $l\in \{ 0,1,...,\nu \}$ and each $m\in \{ 0,1,...,l-1\}$ we have
\begin{equation}  
\Psi _{l,m}(x)=
\frac{{\mathcal A}_m^+ }{\lambda _l-\lambda _m}
\frac{{\mathcal A}_{m+1}^+ }{\lambda _l-\lambda _{m+1}}...
\frac{{\mathcal A}_{l-2}^+ }{\lambda _l-\lambda _{l-2}}
\frac{{\mathcal A}_{l-1}^+ }{\lambda _l-\lambda _{l-1}}
\Psi _{l,l}(x).
\end{equation}  

If we choose the change of variable $s=s(x)$ such that 
$ds/dx=-\kappa (s(x))$, then the formulae 
(\ref{tildeA+}), (\ref{Wm}), (\ref{Vm}) and (\ref{WV}) become
\begin{equation}
{\cal A}_m=-\frac{d}{dx}+W_m(x)\qquad 
{\cal A}_m^+ =\frac{d}{dx}+W_m(x)
\end{equation}
\begin{equation}
W_m(x)=-\frac{\tau (s(x))}{2\kappa (s(x))}
+\frac{2m-1}{2\kappa (s(x))}\frac{d}{dx}\kappa (s(x))
\end{equation}
\begin{equation}
\fl V_m(x)-\lambda _m=W_m^2(x)+\dot W_m(x)\qquad 
V_{m+1}(x)-\lambda _m=W_m^2(x)-\dot W_m(x)
\end{equation}
\begin{equation}
W_m(x)=\frac{\dot \Psi _{m,m}(x)}{\Psi _{m,m}(x)}\qquad \qquad
V_m(x)=\frac{\ddot \Psi _{m,m}(x)}{\Psi _{m,m}(x)}+\lambda _m 
\end{equation}
respectively.\\[3mm]
{\bf Examples.} Let $\alpha _m=(1-\alpha -2m)/2$, \ 
$\alpha '_m=(-1-\alpha +2m)/2$ and $\delta =-\beta /2$.\\
1. In the case $\sigma (s)=1-s^2$, \ $\tau (s)=\alpha s+\beta $, \ $s(x)={\cos x}$ we get
\[ \fl \begin{array}{l}
W_m(x)=\alpha '_m {\cotan }x+\delta {\cosec }x
=\frac{\alpha '_m+\beta }{2}{\cotan }\frac{x}{2}
-\frac{\alpha '_m-\beta }{2}{\tan }\frac{x}{2}\\
V_m(x)=\left( {\alpha '_m}^2-\alpha '_m+\delta ^2 \right) {\cosec }^2x
+ (2\alpha _m-1)\delta {\cotan} x {\cosec }x-{\alpha '_m}^2+\lambda _m\\
{\mbox{}\qquad \qquad  (Poschl-Teller\ type\ potential)}.
\end{array}\]
2. In the case $\sigma (s)=s^2-1$, \ $\tau (s)=\alpha s+\beta $, \ $s(x)={\cosh }\, x$ we get
\[ \fl \begin{array}{l}
W_m(x)=\alpha _m {\cotanh }\, x+\delta {\cosech }x\\
V_m(x)=\left( \alpha _m^2+\alpha _m+\delta ^2 \right) {\cosech }^2x
+ (2\alpha _m+1)\delta {\cotanh }\, x {\cosech }x+\alpha _m^2+\lambda _m\\ 
{\mbox{}\qquad \qquad  (generalized\ Poschl-Teller\ potential)}.
\end{array}\]
3. In the case  $\sigma (s)=s^2$, \ $\tau (s)=\alpha s+\beta $, \ $s(x)=\e^x$ we get
\[ \fl \begin{array}{l}
W_m(x)=\alpha _m+\delta \e^{-x}\\
V_m(x)=\delta ^2\e^{-2x}+(2\alpha _m+1)\delta \e^{-x}+\alpha _m^2+\lambda _m\\ 
{\mbox{}\qquad \qquad (Morse\ type\ potential)}.
\end{array} \]
4. In the case $\sigma (s)=s^2+1$, \ $\tau (s)=\alpha s+\beta$, \  $s(x)={\sinh }\, x$ we get
\[ \fl \begin{array}{l}
W_m(x)=\alpha _m {\tanh }\, x+\delta {\sech }x \\
V_m(x) =\left(-\alpha _m^2-\alpha _m+\delta ^2\right) {\sech }^2x+
(2\alpha _m+1)\delta {\tanh }\, x {\sech }x +\alpha _m^2+\lambda _m\\
{\mbox{}\qquad \qquad (Scarf\ hyperbolic\ type\ potential)}.
\end{array}\]

\section{Concluding remarks}

The equation (\ref{hypeq}) defines an infinite system of orthogonal
polynomials in the first three cases presented in table 1, and a 
finite system of orthogonal polynomials in the last three cases.
Despite the fact that all these orthogonal polynomials can be expressed 
in terms of the classical ones (theorem 2), 
we think that it is worth considering them.
They are directly related to the bound-state eigenfunctions of some 
important Schr\" odinger equations and allow us to analyse these 
equations together, in a unified formalism.
The relation with classical polynomials (\ref{classical}) has
not a very simple form in all the cases. More than that, 
in certain cases we have to consider the classical polynomials outside 
the interval where they are orthogonal or for complex values of parameters. 
Generally, the properties of the functions considered in this paper 
(orthogonality, square integrability,
recursion relations, raising/lowering operators) can not be obtained 
in a simple way from those concerning the classical polynomials.

\ack{This  research was supported by the grant CERES no. 24/2002.}

\end{document}